\magnification=1200
\centerline{IDENTITIES FOR CORRELATION FUNCTIONS IN CLASSICAL}
\centerline{STATISTICAL MECHANICS AND THE PROBLEM OF CRYSTAL STATES.}
\bigskip
\centerline{by David Ruelle\footnote{$\dagger$}{IHES, 91440 Bures sur Yvette, France. email: ruelle@ihes.fr}.}
\bigskip\bigskip\bigskip\bigskip\noindent
	{\leftskip=1.8cm\rightskip=2cm{\sl Abstract:}
\medskip\noindent
Let $z$ be the activity of point particles described by classical equilibrium statistical mechanics in ${\bf R}^\nu$.  The correlation functions $\rho^z(x_1,\dots,x_k)$ denote the probability densities of finding $k$ particles at $x_1,\dots,x_k$.  Letting $\phi^z(x_1,\dots,x_k)$ be the cluster functions corresponding to the $\rho^z(x_1,\dots,x_k)/z^k$ we prove identities of the type
$$	\phi^{z_0+z'}(x_1,\dots,x_k)   $$
$$	=\sum_{n=0}^\infty{z'^n\over n!}\int dx_{k+1}\dots\int dx_{k+n}\,\phi^{z_0}(x_1,\dots,x_{k+n})   $$
It is then non-rigorously argued that, assuming a suitable cluster property (decay of correlations) for a crystal state, the pressure and the translation invariant correlation functions \- $\rho^z(x_1,\dots,x_k)$ are real analytic functions of $z$.\par}
\vfill\eject
	{\sl This note is dedicated to Joel Lebowitz on his 90-th birthday.}
\bigskip
	I first met Joel in the early 1960's.  He was then organizing the famous Yeshiva meetings of statistical mechanics, which later became the Rutgers meetings.  This was a great period for equilibrium statistical mechanics: the old ideas of Maxwell, Boltzmann and Gibbs had uncovered rich mathematical structures which rigorous studies now made progressively more explicit.  Thermodynamic limit, variational principle, analyticity of thermodynamic functions, phase transitions, correlation inequalities, etc., were analyzed and deep results obtained.  The important problems which were left became progressively more and more difficult to attack, and statistical mechanicians turned progressively to problems other than equilibrium.  Here I would like to go back to an old unsolved problem, that of crystals at nonzero temperature.
\medskip
	At the origin of the present note are some considerations on the crystalline state of matter as idealized by the equilibrium statistical mechanics of a classical system of point particles in ${\bf R}^\nu$.  In this description we assume translationally invariant interactions between particles but not necessarily rotational invariance.  In dimension $\nu\ge3$ we expect that there exist crystal Gibbs states where the translation invariance is broken, although no example is known where this has been rigorously proved.  [In dimension $\nu\le2$ the Mermin-Wagner theorem excludes long-range order for short range interactions.]
\medskip
	For suitable particle interactions, we do not see an obvious reason for singularities to occur in the dependence of the crystal equilibrium state on the activity $z$.  We shall argue that there is in fact a real analytic dependence of the pressure and the translationally invariant crystal state on $z$ when we assume a certain cluster property (finite correlation length).  This is not at all a rigorous proof since it depends on unproved technical assumptions.  An important ingredient of the argument can be proved precisely: this ingredient is a relation between the cluster functions $\phi^z$ at activities $z=z_0$ and $z=z_0+z'$ (see (2.9), (2.13), (4.6) below).
\medskip
	In what follows we start with a rigorous study of the $z$-dependance of $\phi^z$ (Sections 1 and 2, which are largely based on Section 4.4 of [5]) and then we proceed with crystal states and a non-rigorous discussion of their cluster properties (Sections 3 and 4).
\medskip\noindent
{\bf 1. Mathematical setup.}
\medskip
	Let ${\cal A}$ be the space of sequences $\phi=(\phi_n)_{n\ge0}$ of bounded complex Lebesgue measurable symmetric functions $\phi_n(x_1,\dots,x_n)$ with $x_1,\dots,x_n\in{\bf R}^\nu$.  This is an Abelian algebra for the product $*$ defined by
$$	(\phi^{(1)}*\phi^{(2)})(X)=\sum_{Y\subset X}\phi^{(1)}(Y).\phi^{(2)}(X\backslash Y)\eqno{(1.1)}   $$
where $X=(x_1,\dots,x_n)$.  (We allow a certain notational confusion between the sequence $(x_1,\dots,x_n)$ and the set $\{x_1,\dots,x_n\}$).  If $\phi\in{\cal A}$, $x\in{\bf R}^\nu$ we define $D_x\phi\in{\cal A}$ by
$$	D_x\phi(x_1,\dots,x_n)=\phi(x,x_1,\dots,x_n)\eqno{(1.2)}   $$
so that
$$	D_x(\phi^{(1)}*\phi^{(2)})=(D_x\phi^{(1)})*\phi^{(2)}+\phi^{(1)}*(D_x\phi^{(2)})\eqno{(1.3)}   $$
i.e., $D_x$ is a derivation on ${\cal A}$.  We also write
$$	D_X\phi=D_{x_1}\dots D_{x_n}\phi\qquad{\rm if}\qquad X=(x_1,\dots,x_n)\eqno{(1.4)}   $$
\medskip
	Let $\Gamma=\exp_*$ denote the exponential for the product $*$.  If we assume $\phi_0=0$ we obtain $(\Gamma\phi)_0=1$ and $(\Gamma\phi)(X)$ is the sum of products $\phi(Y_1)\dots\phi(Y_r)$ over all partitions of $X$ into subsequences $Y_1,\dots,Y_r$ if $X\ne\emptyset$:
$$	\phi_0=0\qquad\Rightarrow\qquad(\Gamma\phi)(X)
	=\sum_{Y_1\sqcup\cdots\sqcup Y_r=X}\phi(Y_1)\dots\phi(Y_r)\eqno{(1.5)}   $$
\medskip
	If $\chi\in L^1({\bf R}^\nu)$, $\phi\in{\cal A}$, and $z$ is a complex variable, we define the formal power series
$$	\langle\chi,\phi\rangle(z)
	=\sum_{n=0}^\infty{z^n\over n!}\int\chi(x_1)dx_1\cdots\int\chi(x_n)dx_n\,\phi_n(x_1,\dots,x_n)\eqno{(1.6)}   $$
obtaining
$$	\langle\chi,\phi^{(1)}*\phi^{(2)}\rangle(z)=\langle\chi,\phi^{(1)}\rangle(z).\langle\chi,\phi^{(2)}\rangle(z)\eqno{(1.7)}  $$
We also have
$$	\langle\chi,\phi\rangle(z_0+z')=\sum_{n=0}^\infty{\sum_{\ell=0}^n{z'^\ell\over \ell!}}{z_0^{n-\ell}\over(n-\ell)!}
 	\int\chi(x_1)dx_1\cdots\int\chi(x_n)dx_n\,\phi_n(x_1,\dots,x_n)   $$
$$	=\sum_{n=0}^\infty{\sum_{\ell=0}^\infty{z'^n\over n!}}{z_0^\ell\over \ell!}
	\int\chi(x_1)dx_1\cdots\int\chi(x_{n+\ell})dx_{n+\ell}\,\phi_{n+\ell}(x_1,\dots,x_{n+\ell})   $$
$$	=\sum_{n=0}^\infty{z'^n\over n!}\int\chi(x_1)dx_1\cdots\int\chi(x_n)dx_n
	\langle\chi,D_{(x_1,\dots,x_n)}\phi)\rangle(z_0)\eqno{(1.8)}   $$
\medskip\noindent
{\bf 2. Statistical mechanics.}
\medskip
	Let
$$Z_\chi=1+\sum_{n>0}{z^n\over n!}\int\chi(x_1)dx_1\cdots\int\chi(x_n)dx_n\,\exp[-\beta U_n(x_1,\dots,x_n)]\eqno{(2.1)}$$
where
$$	z\ge0\quad,\quad\chi\ge0\quad,\quad\beta>0\quad,\quad U_n\ge-nB\quad{\rm with}\quad 0\le B<\infty   $$
and $U_n(x_1,\dots,x_n)$ is symmetric Lebesgue measurable in $x_1,\dots,x_n\in{\bf R}^\nu$, its values being in ${\bf R}\cup\{+\infty\}$.  We also assume $U_0(\emptyset)=U_1(x_1)=0$.  The probability density of finding particles at $x_1,\dots,x_k$ is defined to be
$$	\rho_k^z(x_1,\dots,x_k) =z^k\psi_k^z(x_1,\dots,x_k)\eqno{(2.2)}   $$
where we have introduced the following power series in $z$:
$$	\psi_k^z(x_1,\dots,x_k)   $$
$$	=(Z_\chi)^{-1}\sum_{n=0}^\infty{z^n\over n!}\int\chi(x_{k+1})dx_{k+1}\cdots\int\chi(x_{k+n})dx_{k+n}\,
	\exp[-\beta U_n(x_1,\dots,x_{k+n})]\eqno{(2.3)}   $$
This power series converges near 0 and extends to a meromorphic function on ${\bf C}$.  We also let $\psi_0^z=1$.  Writing $\psi^z=(\psi_k^z)_{k\ge0}$ we have then
$$	\psi^0(X)=\exp[-\beta U(X)]\quad,\quad Z_\chi=\langle\chi,\psi^0\rangle(z)\eqno{(2.4)}   $$
Note that $|Z_\chi|\le\exp[|z|.||\chi||_1e^{\beta B}]$.  From $(2.3), (1.4), (1.6), (1.7), (2.4)$ we obtain
$$   \psi^z(X)=(Z_\chi)^{-1}\langle\chi,D_X\psi^0\rangle(z)
	=(Z_\chi)^{-1}\langle\chi,\psi^0*(\psi^0)^{-1}D_X\psi^0\rangle(z)  $$
$$	=\langle\chi,(\psi^0)^{-1}D_X\psi^0\rangle(z)\eqno(2.5)   $$
\medskip
	Let us define $\phi^z$ by
$$	\psi^z=\Gamma\phi^z\eqno{(2.6)}   $$
so that $\phi_0^z=0$.  From $(2.5), (2.6), (1.7), (1.5)$ we obtain
$$	\psi^z(X)=\langle\chi,(\Gamma\phi^0)^{-1}D_X\Gamma\phi^0\rangle(z)=\sum_{Y_1\sqcup\cdots\sqcup Y_r=X}
	\langle\chi,D_{Y_1}\phi^0*\dots*D_{Y_r}\phi^0\rangle(z)   $$
$$	=\sum_{Y_1\sqcup\cdots\sqcup Y_r=X}\langle\chi,D_{Y_1}\phi^0\rangle(z)\dots
	\langle\chi,D_{Y_r}\phi^0\rangle(z)=(\Gamma[\langle\chi,D_{.}\phi^0\rangle(z)])(X)\eqno{(2.7)}   $$
so that (2.6), (2.7), (1.6) yield
$$	\phi_k^z(x_1,\dots,x_k)=\langle\chi,D_{(x_1,\dots,x_k)}\phi^0\rangle(z)   $$
$$	=\sum_{n=0}^\infty{z^n\over n!}\int\chi(x_{k+1})dx_{k+1}\cdots\int\chi(x_{k+n})dx_{k+n}\,
	\phi^0_n(x_1,\dots,x_{k+n})\eqno{(2.8)}$$
\medskip
	Using repeatedly (2.8) and (1.8) we obtain
$$	\phi_k^{z_0+z'}(x_1,\dots,x_k)=\langle\chi,D_{(x_1,\dots,x_k)}\phi^0\rangle(z_0+z')   $$
$$	=\sum_{n=0}^\infty{z'^n \over n!}\int\chi(x_{k+1})dx_{k+1}\cdots\int\chi(x_{k+n})dx_{k+n}
	\langle\chi,D_{(x_{k+1},\dots,x_{k+n})}D_{(x_1,\dots,x_k)}\phi^0)\rangle(z_0)   $$
$$	=\sum_{n=0}^\infty{z'^n\over n!}\int\chi(x_{k+1})dx_{k+1}\cdots\int\chi(x_{k+n})dx_{k+n}\,
	\phi_{k+n}^{z_0}\,(x_1,\dots,x_{k+n})\eqno{(2.9)}  $$
\indent
	We note also that the $\phi_k^z$ can be expressed as functional derivatives of $\log Z_\chi$ as in (2.10) below.  Let us define
$$	\tilde\phi_k^z(x_1,\dots,x_k)={\delta^k\log Z_\chi\over\delta\chi^k}(x_1,\dots,x_k)
	={\delta[\tilde\phi_{n-1}^z(x_1,\dots,x_{n-1})]\over\delta\chi}(x_n)   $$
Then
$$	{\delta Z_\chi\over\delta\chi}(x_1)=\tilde\phi_1^z(x_1).Z_\chi   $$
hence using (2.3)
$$	\psi^z(X).Z_\chi={\delta^{|X|}Z_\chi\over\delta\chi^{|X|}}(X)
	=\sum_{Y_1\sqcup\cdots\sqcup Y_r=X}\tilde\phi^z(Y_1)\dots\tilde\phi^z(Y_r).Z_\chi   $$
so that using also (1.5), (2.6), we have $\tilde\phi^z=\phi^z$, i.e.,
$$	\phi_k^z(x_1,\dots,x_k)={\delta^k\log Z_\chi\over\delta\chi^k}(x_1,\dots,x_k)\eqno{(2.10)}   $$
\medskip
	Let ${\cal Z}$ be a subgroup of ${\bf R}^\nu$ with $\nu$ linearly independent generators, and $m$ a positive integer.  We can replace ${\bf R}^\nu$ by the torus ${\bf T}_m^\nu={\bf R}^\nu/m{\cal Z}$ and use a suitable definition of $U(x_1,\dots)$ for $x_1,\dots\in{\bf T}_m^\nu$ (we may assume that the forces defining $U$ to have exponentially short range in some sense, see below, and take $m$ large).  We define then
$$	Z_m=1+\sum_{n>0}{z^n\over n!}\int_{{\bf T}_m^\nu}dx_1\dots\int_{{\bf T}_m^\nu}dx_n\exp[-\beta U(x_1,\dots,x_n)]
	\eqno{(2.11)}   $$
$$	\psi_{m,k}^z(x_1,\dots,x_k)
	=(Z_m)^{-1}\sum_{n=0}^\infty{z^n\over n!}\int_{{\bf T}_m^\nu}dx_{k+1}\dots\int_{{\bf T}_m^\nu}dx_{k+n}
	\exp[-\beta U(x_1,\dots,x_{k+n})]\eqno{(2.12)}   $$
Writing $\psi_m^z=\Gamma\phi_m^z$ we obtain
$$	\phi_{m,k}^{z_0+z'}(x_1,\dots,x_k)=\sum_{n=0}^\infty{z'^n\over n!}\int_{{\bf T}_m^\nu}dx_{k+1}\cdots
	\int_{{\bf T}_m^\nu }dx_{k+n}\,\phi_{m,{k+n}}^{z_0}\,(x_1,\dots,x_{k+n})\eqno{(2.13)}  $$
instead of (2.9).
\medskip\noindent
{\bf 3. Crystal states.}
\medskip
	We place ourselves now in the usual framework of classical equilibrium statistical mechanics.  Under suitable assumptions on the potentials $U_n$, there are infinite volume limits of the $\rho_k^z$ defined by (2.2), (2.3), or (2.12), and these limits correspond to {\it Gibbs states} $\sigma$ (see [2] and later papers, [4]\footnote{*)}{in [4] Gibbs states are called equilibrium states which are not necessarily invariant under translations.}, [6], [1]).  Gibbs states for interactions with hard cores are defined in [4].  The {\it infinite volume} limits are obtained when the function $\chi:{\bf R}^\nu\mapsto[0,1]$ tends to 1 on bounded subsets of ${\bf R}^\nu$, or when $m\to\infty$ in ${\bf T}_m^\nu$.
\medskip
	Note that the pressure $p(z)$ is defined as an infinite volume limit
$$	p(z)=\lim_{\chi\to1}[\beta\int\chi(x)\,dx]^{-1}\log Z_\chi=\lim_{m\to\infty}[\beta\int_{{\bf T}_m^\nu}dx]^{-1}\log Z_m   $$
\indent
	For definiteness we shall assume that the $U_n$ correspond to a translation invariant (but not necessarily rotation invariant) exponentially decreasing interaction with hard cores between particles (see [4], [3]).  There are thus $r>0, R>0$ such that
$$	U_n(X)=\sum_{Y\subset X,|Y|\ge2}\Phi(Y)\eqno{(3.1)}   $$
where $Y\mapsto\Phi(Y)$ is continuous on
$$	{\cal Y}_k=\{Y:|Y|=k,\min(\{|y-x|:x,y\in Y\})\ge r\}   $$
and satisfies
$$	\Phi(Y+x)=\Phi(Y)\;\,{\rm if}\;\,x\in{\bf R}^\nu\;\;,\;\;\Phi(\{x,y\})=+\infty\;\,{\rm if}\;\,|y-x|<r   $$
$$	\lim_{{\rm diam}Y\to\infty}\Phi(Y)\exp({\rm diam}Y/R)=0   $$
The set of these $Y$ is a separable Banach space with the norm
$$	||\Phi||=\sup_{Y\in{\cal Y}_k,k\ge2}|\Phi(Y)|\exp({\rm diam}Y/R)<+\infty   $$
\indent
	Note that the set of Gibbs states corresponding to the infinite volume limit of $\psi^z$ or $\psi_m^z$ is invariant under translations.  Therefore there exists a translation invariant Gibbs state $\sigma$ and this invariant Gibbs state is generically unique.  (We use here [4], [3], and Mazur's theorem: a convex function on a separable Banach space has a unique tangent on a dense $G_\delta$, see [7] p. 291).  Given ${\cal Z}$ the Gibbs state $\sigma_m$ corresponding to $\psi_m^z$ is invariant under translations of ${\bf T}_m^\nu$ and therefore its limits when $m\to\infty$ are translation invariant Gibbs states.
\medskip
	The Gibbs states for given $z$ form a {\it simplex} (see for instance [4]).  Therefore there is a unique decomposition of the translational invariant Gibbs state $\sigma$ into extremal Gibbs states.  We say that $\sigma$ is a {\it crystal state} if there is a subgroup ${\cal Z}\subset{\bf R}^\nu$ such that the extremal Gibbs states into which $\sigma$ is decomposed are invariant under the maximal subgroup ${\cal Z}$, where ${\cal Z}$ has $\nu$ generators and the quotient ${\bf R}^\nu/{\cal Z}$ is compact.  We can then write $\sigma=\int\mu^z(d\tau)\,\tau\hat\sigma^z$ where $\mu^z$ is the Haar measure on ${\bf R}^\nu/{\cal Z}$ and $\hat\sigma^z$ is any of the extremal Gibbs states into which $\sigma$ is decomposed.  If $K^z$ is a fundamental domain for ${\cal Z}$, i.e., ${\bf R}^\nu=\sqcup_{u\in{\cal Z}}(K^z+u)$ we can also write
$$	\psi^z(x_1,\dots,x_k)=\int_{K^z}dv\,\hat\psi^z(x_1+v,\dots,x_k+v)\eqno{(3.2)}   $$
where $dv$ is the normalized Lebesgue measure on $K^z$ and $\hat\psi^z$ corresponds to $\hat\sigma^z$.  We assume from now on that the translationally invariant Gibbs state $\sigma$ is extremal, i.e., ${\bf R}^\nu$-ergodic (see [5] p. 161).  If the $U_n$ are rotationally invariant the ${\bf R}^\nu$-ergodic states correspond to the various possible crystal orientations in ${\bf R}^\nu$.
\medskip\noindent
{\bf 4. Cluster property.}
\medskip
	Let ${\cal Z}$ be the subgroup of ${\bf R}^\nu$ associated with the ${\bf R}^\nu$-ergodic crystal state $\sigma$, and let $\psi_{m,k}^z(x_1,\dots,x_k)$ be the corresponding function of $x_1,\dots,x_k\in{\bf T}_m^\nu$.  We use the notation
$$	\Phi(\hat x_j)=\int_{K^z}dv_j\Phi(x_j+v_j)\eqno{(4.1)}   $$
for the integral of $\Phi$ over $x_j+K^z$.  We assume the following {\it cluster property}:
$$	\psi_{m,{k+n}}^z(X,\hat x_{k+1},\dots,\hat x_{k+n})
-\psi_{m,{k+r}}^z(X,\hat x_{k+1},\dots,\hat x_{k+r})
	\psi_{m,{n-r}}^z(\hat x_{k+r+1},\dots,\hat x_{k+n})\to0\eqno{(4.2)}   $$
for given $X=(x_1,\dots,x_n)$, uniformly in $m$, when
$$	{\rm dist}(\{X,x_{k+1},\dots,x_{k+r}\},\{x_{k+r+1},\dots,x_{k+n}\})\to\infty\eqno{(4.3)}   $$
[The integration over $v_j$ in (4.1) corresponds to the fact that $\sigma$ is expected to be only ${\cal Z}$-mixing, not ${\bf R}^\nu$-mixing].
\medskip
	In the case of a rotationally invariant interaction between the particles, we may hope that the choice of ${\cal Z}$ selects an orientation of the crystal, so that the limit $m\to\infty$ corresponds to an extremal translationally invariant state $\sigma$ and the cluster property is satisfied.
\medskip
	Given $n\ge1$ the cluster property (4.2) will imply that
$$	\phi_{m,{k+n}}^z(X,\hat x_{k+1},\dots,\hat x_{k+n})\to0   \eqno{(4.4)}  $$
uniformly in $m$ when (4.3) holds.  We prove this by induction on $n$.  Since (4.3) implies (4.2) we have, using (1.5),
$$	\psi_{m,{k+n}}^z(X,\hat x_{k+1},\dots,\hat x_{k+n})
	-\psi_{m,{k+r}}^z(X,\hat x_{k+1},\dots,\hat x_{k+r})\psi_{m,{n-r}}^z(\hat x_{k+r+1},\dots,\hat x_{k+n})   $$
$$	=\phi_{m,{k+n}}^z(X,\hat x_{k+1},\dots,\hat x_{k+n})
	+\hbox{asymptotically vanishing products of $\phi$'s}   $$
The asymptotically vanishing terms are obtained from the induction assumption.  Equation (4.4) follows thus from the cluster property (4.2).
\medskip
	Therefore given $\epsilon>0$ there is $K>$ diam$X$ such that
$$	|\phi_{m,k+n}^z(X,\hat x_{k+1},\dots,\hat x_{k+n})|<\epsilon   $$
unless for every pair of elements in the set $\{X,x_{k+1},\dots,x_{k+n}\}$ there is a sequence of elements connecting the pair with distance $<K$ between successive elements, in which case the elements of $\{X,x_{k+1},\dots,x_{k+n}\}$ can be arranged in a tree such that neighboring elements of the tree have distances $<K$.
\medskip
	We use Cayley's formula for counting trees and assume a cluster property for $\psi_{m,k+n}^{z_0}$ characterized by a finite integral $C$ to guess an estimate of the form
$$	\int_{{\bf T}_m^\nu}dx_{k+1}\dots\int_{{\bf T}_m^\nu}dx_{k+n}\,|\phi_{m,{k+n}}^{z_0}(X,x_{k+1},\dots,x_{k+n})|
	\le AC^nn^{n-2}\eqno{(4.5)}   $$
where $A$ depends on $k$.
\medskip
	In the infinite volume limit (2.13) gives
$$	\phi_k^{z_0+z'}(x_1,\dots,x_k)
	=\sum_{n=0}^\infty{z'^n\over n!}\int dx_{k+1}\dots\int dx_{k+n}\,\phi_{k+n}^{z_0}(x_1,\dots,x_{k+n})\eqno{(4.6)}   $$
where (4.5) implies
$$	\Big|{z'^n\over n!}\int dx_{k+1}\dots\int dx_{k+n}\,\phi_{k+n}^{z_0}(x_1,\dots,x_{k+n})\Big|
	\le A(C|z'|)^n{n^{n-2}\over n!}\le A(Ce|z'|)^n\eqno{(4.7)}   $$
Therefore $\phi_k^z(x_1,\dots,x_k)$ is a real analytic function of $z$ for $z$ close to $z_0$.  Hence the same is also true of $\psi_k^z(x_1,\dots,x_k)$.
\medskip
	Note that $\phi_{m,k}^z(x_1,\dots,x_k)$ is translation invariant so that $\phi_{m,1}^z(x_1)=\phi_{m,1}^z$ is constant.  Therefore
$$	{1\over m^\nu}[\log Z_m(z_0+z')-\log Z_m(z_0)]=\int_{z_0}^{z_0+z'}dz\,\phi_{m,1}^z   $$
is real analytic in $z'$ and the same is true of $p(z_0+z')-p(z_0)$.
\medskip
	Let us summarize.  We assume that for an interaction of the type considered and for a given value $z_0$ of the activity there is an ${\bf R}^\nu$-ergodic Gibbs state $\sigma$, that this is a crystal state, and we also assume a cluster property with the specific consequence (4.5).  Then the pressure $p(z)$ and the $\psi_k^z(x_1,\dots,x_k)$ are real analytic functions of $z$ for $z$ close to $z_0$.
\vfill\eject
\noindent
{\bf References.}
\medskip\noindent
[1] S. Dashian and B.S. Nahapetian.  ``On the relationship of energy and probability in models of classical statistical physics.''  Markov Processes Relat. Fields {\bf 25},649-681(2019).
\medskip\noindent
[2] R.L. Dobrushin.  ``The description of a random field by means of conditional probabilities and conditions of its regularity.''  Theory Prob. Applications {\bf 13},197-224(1968).
\medskip\noindent
[3] H.-O. Georgii.  ``Large deviations for hard-core particle systems'' pp. 108-116 in R. Koteck\'y. {\it Phase Transitions}, World Scientific, Singapore, 1993.
\medskip\noindent
[4] O.E. Lanford and D. Ruelle ``Observables at infinity and states with short range correlations in statistical mechanics.''  Commun. Math. Phys. {\bf 13},194-215(1969).
\medskip\noindent
[5] D. Ruelle.  {\it Statistical Mechanics, Rigorous Results}.  Benjamin, New York, 1969.
\medskip\noindent
[6] D. Ruelle.  {\it Thermodynamic Formalism}.  Addison-Wesley, Reading MA, 1978.
\medskip\noindent
[7] B. Simon.  {\it Convexity: an analytic viewpoint.}  Cambridge U.P., 2011.
\end